\documentclass[10pt,conference]{IEEEtran}
\IEEEoverridecommandlockouts
\usepackage{cite}
\usepackage{amsmath,amssymb,amsfonts}
\usepackage{algorithmic}
\usepackage{graphicx}
\usepackage{textcomp}
\usepackage{xcolor}

\usepackage[multiple]{footmisc}
\def\BibTeX{{\rm B\kern-.05em{\sc i\kern-.025em b}\kern-.08em
    T\kern-.1667em\lower.7ex\hbox{E}\kern-.125emX}}

\usepackage{enumitem}
\usepackage{tcolorbox}
\usepackage{mdframed}
\usepackage{amsmath}

\begin{document}
\bstctlcite{IEEEexample:BSTcontrol}

\title{
Examining the Relationship of Code and Architectural Smells with Software Vulnerabilities
}

\author{\IEEEauthorblockN{Kazi Zakia Sultana}
\IEEEauthorblockA{
\textit{Montclair State University}\\
sultanak@montclair.edu}
\and
\IEEEauthorblockN{Zadia Codabux}
\IEEEauthorblockA{
\textit{University of Saskatchewan}\\
zadiacodabux@ieee.org
}
\and
\IEEEauthorblockN{Byron Williams}
\IEEEauthorblockA{
\textit{University of Florida}\\
byron@cise.ufl.edu
}
}
\maketitle
\begin{abstract}
\textit{Context:} Security is vital to software developed for commercial or personal use. Although more organizations are realizing the importance of applying secure coding practices, in many of them, security concerns are not known or addressed until a security failure occurs. The root cause of security failures is vulnerable code. While metrics have been used to predict software vulnerabilities, we explore the relationship between code and architectural smells with security weaknesses. As smells are surface indicators of a deeper problem in software, determining the relationship between smells and software vulnerabilities can play a significant role in vulnerability prediction models. \textit{Objective:} This study explores the relationship between smells and software vulnerabilities to identify the smells. \textit{Method:} We extracted the class, method, file, and package level smells for three systems: Apache Tomcat, Apache CXF, and Android. We then compared their occurrences in the vulnerable classes which were reported to contain vulnerable code and in the neutral classes (non-vulnerable classes where no vulnerability had yet been reported). \textit{Results:} We found that a vulnerable class is more likely to have certain smells compared to a non-vulnerable class. \textit{God Class, Complex Class, Large Class, Data Class, Feature Envy, Brain Class} have a statistically significant relationship with software vulnerabilities. We found no significant relationship between architectural smells and software vulnerabilities. \textit{Conclusion:} We can conclude that for all the systems examined, there is a statistically significant correlation between software vulnerabilities and some smells.\\
\end{abstract}

\begin{IEEEkeywords}
Vulnerability, Code Smell, Architectural Smell, Software Security
\end{IEEEkeywords}

\section{Introduction}
Software is critical to many industries. Software development organizations need to ensure that software safeguards sensitive and confidential information while providing quality services to its users. Software companies must fulfill the requirements of their stakeholders while preventing malicious attempts from potential attackers. In addition to security measures like firewalls and intrusion detection systems, designing and developing secure software should also be of paramount importance. However, in many software companies, security concerns are often not known or addressed early in the software development lifecycle and only become known when they manifest themselves as post-release vulnerabilities.

A vulnerability is ``a security exposure that results from a product weakness
that the product developer did not intend to introduce and should fix once it is
discovered''.\footnote{https://msdn.microsoft.com/en-us/library/cc751383.aspx} MITRE Corporation defines an information security vulnerability as ``a mistake in
software that can be directly used by a hacker to gain access to a system or
network''.\footnote{https://cve.mitre.org/about/terminology.html} For instance, injection flaws (e.g., SQL injection) occur when untrusted data is sent to an interpreter as part of a command or query.\footnote{https://www.owasp.org/index.php/Top\_10\_2013-Top\_10} While traditional metrics have been used to detect software vulnerabilities, we intend to uncover additional ways of identifying software vulnerabilities. In this study, we aim to determine whether there is a relationship between code and architectural smells (referred to as ``smells'' in this paper) and software vulnerabilities and to determine what are the leading smells in vulnerable classes. 
Companies such as Facebook and Google that have rigorous quality processes in place are already extracting smells or performing static analysis ~\cite{distefano2019scaling,sadowski2018lessons}. Therefore, software practitioners can make additional use of this data to identify vulnerable components in their software. 

Code smell, a term coined by Kent Beck, refers to a surface indication that usually corresponds to a deeper problem in the system~\cite{Fowler}. However, it may not be true in all cases as advocated by Martin Fowler: ``Smells aren't inherently bad on their own - they are often an indicator of a problem rather than the problem themselves''.\footnote{https://martinfowler.com/bliki/CodeSmell.html} Many studies have identified code smells as indicators of fault-proneness that directly relate to the system's quality~\cite{Khomh, Li, Yamashita}. Code smells are also related to weak system design, bad implementation methods, micro and nano-patterns which may lead to poor quality~\cite{Tufano, codabux2017, codabux2017relationship}. So, the identification and management of smells is an important part of software quality assurance to build quality software.

Although researchers investigated the correlations between vulnerabilities and code smells~\cite{Islam2017, Elkhail, Ghafari, Mumtaz}, a more extensive and rigorous study on analyzing code smells and their impact on vulnerabilities is still needed for secure software development. Most of the current research focused on a particular context, for example, either they worked on a specific class of code smells to relate them to vulnerabilities or they focused on the vulnerabilities in a particular system like Android. Besides, to the best of our knowledge, this is the first study on investigating the relationship between architecture smells (defined later) and software vulnerabilities.

Ten common defects account for about 75\% of all software vulnerabilities~\cite{Conklin}. Existing studies are mostly confined to predicting software vulnerabilities using software metrics (e.g., code churn, complexity, fault history)~\cite{Chowdhury:2011, Zimmermann}. Although traditional software defect metrics can be used for predicting the existence of vulnerabilities, vulnerability prediction is a
bit challenging since vulnerability data is relatively limited, resulting in difficulties for training a model~\cite{YShin}. Another study that evaluated 45 e-business applications showed that 70\% of security defects were caused by poor software design~\cite{Jaquith}. According to Mumtaz et al.~\cite{Mumtaz}, ``Code smells indicate design flaws that can degrade the quality of software and can potentially lead to the introduction of faults.'' Therefore, the objective of this work is to explore whether smells extracted from source code are related to software vulnerabilities.  


The contributions of this study are: 
\begin{itemize}[noitemsep,topsep=0pt]
    \item  We analyze the smells in the vulnerable and neutral (non-vulnerable) code of the systems under study to find the correlation between software vulnerabilities and smells. The correlation analysis will help guide software practitioners in secure software development by indicating area in the codebase that are more prone to vulnerability. For example, as developers are made aware of certain vulnerable smells, they can modify their testing strategy and apply appropriate quality assurance approaches (e.g., static analysis).
    
    \item This study can be used as a baseline to establish a prediction model for software vulnerabilities using smells. Smell extraction is achieved using a set of metrics and thresholds that have been validated by existing research~\cite{Munro, Lanza}. This study examining the relationship between smells and software vulnerabilities opens a new research direction of vulnerability prediction using smells.  
\end{itemize}

The rest of the paper is organized as follows. Section~\ref{background} introduces background information on code and architecture smells. Section~\ref{related} highlights the relevant related work. Section~\ref{method} focuses on the methodology of the study including the research goal, questions, and study design. Section~\ref{results} describes the data analysis and provides insights into the results. Section~\ref{discuss} presents the discussion. Section~\ref{threats} lists the threats to the validity of this study. Section~\ref{conc} concludes and outlines future work.

\section{Background}
\label{background}

In this study, we refer to a class as a \textit{vulnerable class} if it has been confirmed and formally reported to contain vulnerable code by a developer (e.g., a task that results in a defect being fixed). A \textit{neutral class} is a class where no known vulnerability has yet been reported. The scope of this work is restricted to determining the correlation (not causation) between vulnerabilities and smells.

Code or bad smells are potential design weaknesses in source code that can cause the software to be more change and defect prone, slow down the software development process and make software maintenance harder~\cite{olbrich2010, Khomh, Yamashita}. 
For example, a God Class (GC) code smell is a central class in a system that has taken on a lot of responsibilities, delegating only trivial tasks to other classes. In short, a god class is a class ``which knows too much and does too much.'' It violates the Single Responsibility Principle. A god class features a high complexity, low cohesion, and heavy access to data of foreign classes and is defined using the following formula~\cite{Lanza}: 
\\
\noindent
\scalebox{0.8}{
$GC = (AFTD > FEW) \land (WMC >= VERY HIGH) \land (TCC < 0.33)$}

where Access to Foreign Data (ATFD) represents the number of external classes from which a given class accesses attributes, directly or via accessor-methods; Weighted Method Count (WMC)) is the sum of the cyclomatic complexity of all methods in a class; Tight Class Cohesion (TCC) is the relative number of methods directly connected via attributes.

Code smells are also referred to as design smells and the terminologies ``code smell'' and ``design smell'' have been used interchangeably in the literature~\cite{palma2015study, alkharabsheh2019software}. However, design smells are also perceived as being distinct from code smells. A design smell is defined as a weakness in the design that violates fundamental design principles and negatively impacts quality~\cite{suryanarayana2014refactoring}. A code smell is also considered as an indicator or symptom of design smell~\cite{moha2007decor}.

Architecture or architectural smells, as introduced by Garcia et al.~\cite{garcia2009toward}, are ``decisions that negatively impact system internal quality. Architectural smells may be caused by applying a design solution in an inappropriate context, mixing design fragments that have undesirable emergent behaviors, or applying design abstractions at the wrong level of granularity.'' Architectural smells are defined at the architecture (higher) level (e.g., components, connectors, styles, packages, subsystems, communications). In our study, we considered the smells: \textit{Hub-Like Dependency, Unstable Dependency, Cyclic Dependency, and Unhealthy Inheritance Hierarchy} that Fontana et al.~\cite{Fontana2019} categorized as architectural smells as they indicate architectural decisions that negatively impact the system's internal quality. 
Some code smells have been classified as both code smell or design smell e.g. god class is referred to as a code smell~\cite{olbrich2010}, design smell~\cite{vaucher2009tracking} and even architecture smell~\cite{azadi2019}. 

We adopt the convention that code/design smells are too similar to distinguish. Therefore, we evaluate two types of smells in our study: code/design smells and architectural smells. For simplicity, smells that assume both the code smell and design smell characteristics will be referred to as ``code smells.'' 
For this study, we extracted smells at class, method, package, and file levels. The smells considered in this study are described in Table~\ref{codesmell}. 

\begin{table*}[!htb]
\begin{minipage}{\linewidth} 
\caption{Smells Considered in the Study}
\label{codesmell}
\centering
\scalebox{0.8}{
\begin{tabular}{llll}
\hline
\textbf{Granularity} & 
\textbf{Smell Type} & 
\begin{tabular}[t]{@{}l@{}l@{\qquad}|}
\textbf{Smell}
\end{tabular} & 
\begin{tabular}[t]{@{}l@{}l@{\qquad}|}
\textbf{Description} 
\end{tabular} \\ 

\hline
Class 
& Code & God Class & \begin{tabular}[t]{@{}l@{}l@{\qquad}|} A class that tends to centralize the intelligence of a system, performs most of the work, delegating only minor \\ details  to a set of trivial classes and using the data from other classes ~\cite{Lanza}
 \end{tabular}\\

& Code & Lazy Class & \begin{tabular}[t]{@{}l@{}l@{\qquad}|} A class that is not doing enough. It either needs to be removed or its responsibility needs to increase ~\cite{mantyla2004} \end{tabular}\\

& Code &  Complex Class & \begin{tabular}[t]{@{}l@{}l@{\qquad}|} A class having at least one method having a high cyclomatic complexity ~\cite{bavota2015experimental} \end{tabular}\\

& Code &  Large Class & \begin{tabular}[t]{@{}l@{}l@{\qquad}|} A class that has a large number of lines of code, global variables, and methods, as well as a complicated \\ function~\cite{Fowler} 
 \end{tabular}\\

& Code &  Data Class & \begin{tabular}[t]{@{}l@{}l@{\qquad}|} A dumb data holder without complex functionality but other classes strongly rely on them~\cite{Lanza}  \end{tabular}\\

& Code &  Refused Bequest & \begin{tabular}[t]{@{}l@{}l@{\qquad}|} A subclass that has a weak connection to its parent's class, by insufficiently using the data and methods \\ inherited from the parent's class \cite{Lanza}
 \end{tabular}\\

& Code &  Brain Class & \begin{tabular}[t]{@{}l@{}l@{\qquad}|} A class that tends to centralize the functionality of a system but unlike god classes, they do not use too much \\ data  from foreign classes and are a little more cohesive \cite{Lanza}
 \end{tabular}\\

& Architecture &  Hub-Like Dependency &  \begin{tabular}[t]{@{}l@{}l@{\qquad}|} When an abstraction has (outgoing and incoming) dependencies on a large number of other abstractions ~\cite{fontana2017arcan}
 \end{tabular}\\

\hline

Method & Code & Feature Envy & \begin{tabular}[t]{@{}l@{}l@{\qquad}|} A method that seems more interested in the data of other classes than that of their own class ~\cite{Lanza}\end{tabular}\\

& Code &  Long Method & \begin{tabular}[t]{@{}l@{}l@{\qquad}|} A method that is too long and tries to do too much on its own~\cite{mantyla2004}\end{tabular}\\

& Code &  Long Parameter List & \begin{tabular}[t]{@{}l@{}l@{\qquad}|} A method that takes too many parameters~\cite{mantyla2004} \end{tabular}\\

& Code &  Brain Method & \begin{tabular}[t]{@{}l@{}l@{\qquad}|} A method that tends to centralize the functionality of a class~\cite{Lanza} 
 \end{tabular}\\

& Code &  Shotgun Surgery & \begin{tabular}[t]{@{}l@{}l@{\qquad}|}  When the method changes, it implies many small changes to a lot of different classes~\cite{Lanza} \end{tabular}\\

\hline

Package & Architecture & Unstable Dependency & \begin{tabular}[t]{@{}l@{}l@{\qquad}|}  A subsystem (component) that depends on other subsystems that are less stable than itself, with a possible \\ ripple effect of changes in the project~\cite{fontana2017arcan} 
 \end{tabular}\\

& Architecture &  Cyclic Dependency\begin{tiny}\footnote{Cyclic Dependency can be at both class or package level granularity} \end{tiny}
& \begin{tabular}[t]{@{}l@{}l@{\qquad}|}  A subsystem (component) that is involved in a chain of relations that break the desirable acyclic nature of a \\ subsystem’s dependency  structure~\cite{fontana2017arcan}
 \end{tabular}\\

\hline

File & Architecture & Unhealthy Inheritance Hierarchy & \begin{tabular}[t]{@{}l@{}l@{\qquad}|} A parent class that depends on its children, or a client class depending on both parent and child~\cite{Mo}  \end{tabular}\\
\hline
\end{tabular}
}
\end{minipage}
\end{table*}

\section{Related Work}
\label{related}


In this section, we highlight existing research that focused on the relationship between smells and vulnerabilities.    
Elkhail et al.~\cite{Elkhail} conducted an empirical study to analyze the relationship between code smells and vulnerabilities. They used Apache Tomcat software vulnerabilities which were identified by static code analysis. According to their study, most code smells (except ``Field Declarations Not at Start of Class'') have a slight impact on vulnerabilities. Islam et al.~\cite{Islam2016, Islam2017} performed a large quantitative empirical
study of the software vulnerabilities in cloned and non-cloned code for C and Java-based open-source software systems. Code clones are the stinkiest of all of the code smells~\cite{Fowler}. They identified a set of five vulnerabilities that appear
more frequently in cloned code compared to non-cloned code~\cite{Islam2016}. They also revealed that the software vulnerabilities found in code clones have higher severity of security risks compared to those in non-cloned code~\cite{Islam2017}. Mumtaz et al.~\cite{Mumtaz} showed that refactoring of bad smells helps improve the security of an application without compromising the overall quality of software systems. 

Ghafari et al.~\cite{Ghafari} introduced the term ``security smells'' - symptoms in the code that signal the prospect of a security vulnerability. They identified 28 smells that may lead to vulnerabilities in Android-powered devices. They concluded that the identified security smells are a good indicator of
software vulnerabilities~\cite{Ghafari}. Gadient et al.~\cite{gadient2019} extended the work to focus on identifying security smells in Android Inter-Component Communication (ICC) vulnerabilities which is one of the Android's most widespread security issues. Rahman et al.~\cite{rahman2019} investigated security smells in the context of Infrastructure as Code (IaC) scripts. They extracted the smells using a static analysis tool, then mapped the smells to security weaknesses using the CWE database.\footnote{https://https://cwe.mitre.org/index.html} They reported that some smells have a long lifetime and provided some guidelines to practitioners for secure IaC script writing. The authors extended their empirical study to include nine security smells for Ansible and Chef IaC scripts~\cite{rahman2019}. Rahman et al.~\cite{rahman2019share} also conducted a similar analysis of the GitHub Gist written in Python. They extracted 13 smells using static analysis before performing an empirical study to identify security smells. 

Most of these studies are conducted in specific domains (e.g. IaC Scripting, Python Gists) and their main focus is security smells. Our study is more general in that we extract a wider range of smells encountered by the developers. A focus on security smells only limits the applicability and may miss seemingly innocuous smells that occur regularly in source code but also strongly correlate to vulnerabilities. 

\section{Methodology}
\label{method}
\subsection{Research Questions}
The goal of this study is \textit{to investigate the relationship between the presence of smells and software vulnerabilities.} This goal will be addressed by the following research questions.

\noindent \emph{ \textbf{RQ1: What is the relationship between smells and software vulnerabilities?}}

We investigate whether classes with smells are more prone to software vulnerabilities compared to others. This information can help indicate potential security flaws and risky areas in software systems. Software companies striving to secure their software can use such information as a proactive measure to identify and prevent potential security breaches which can be financially detrimental to the company. 

If we show that code smells present different distributions in two groups: vulnerable vs. neutral classes, we can infer that the presence of smells is related to vulnerabilities. This finding can lead to further investigation of the type of relationship among them. 
We formulate the following hypotheses:



\noindent \textbf{Null Hypothesis ($H_{01}$):} \emph{Smells are not associated with software vulnerabilities.}

\noindent \textbf{Alternative Hypothesis ($H_{A1}$):} \emph{Smells are more likely to be associated with software vulnerabilities.}

\noindent \emph{ \textbf{RQ2: Are there certain smells that have a stronger relationship to vulnerabilities than others?}}

We investigate whether classes with specific kinds of smells contribute more to software vulnerabilities compared to others. This analysis will help us find the problematic types of smells in vulnerable and neutral code. For example, if we find that \textit{Large Class} has a significantly different percentage of occurrences in vulnerable classes compared to the neutral classes, we can identify \textit{Large Class} as having a relationship with vulnerability. In that case, developers will be able to identify and address \textit{Large Class} in the code. On the other hand, if \textit{Large Class} smell does not show significant differences in their presence in vulnerable vs. neutral classes, we can conclude that smell has no relationship or a weak relationship with vulnerability. Accordingly, we formulated the hypotheses:

\noindent \textbf{Null Hypothesis ($H_{02}$):} \emph{A particular type of smell in a class is not associated with a software vulnerability.}

\noindent \textbf{Alternative Hypothesis ($H_{A2}$):} \emph{A particular type of smell in a class is more likely to be associated with a software vulnerability.}

To establish a strong correlation between smells and vulnerabilities, there should be a significant difference in the number of smells in vulnerable classes compared to neutral classes.

\subsection{Study Design}
\label{design}
The study consists of the smells and software vulnerabilities captured from three systems: Apache Tomcat, Apache CXF, and Android. 

The rationale behind choosing these systems are as follows:
\begin{itemize}[noitemsep,topsep=0pt]
    \item The systems have well-documented public vulnerability repositories.
    \item The systems are open-source and developed for different types of web services.
    \item Our tool detects smells in Java code and the tool is applicable for these systems as they are fully or mostly developed in Java. 
    \item The systems are heavily used for building and deploying real-world applications.
\end{itemize}

We analyzed 37 versions of Apache Tomcat, 17 versions of Apache CXF, and 2 versions of Android. The characteristics of the analyzed systems are presented in Table~\ref{stats}. The number of classes, methods, and lines of code (LOC) are based on the last version of each major release at the time of the study.

\begin{table}[!htb]
\caption{Project Statistics}
\label{stats}
\centering
\scalebox{0.7}{
\begin{tabular}{lclccc}
\hline
\textbf{Systems}
& \begin{tabular}[t]{@{}l@{}l@{\qquad}|}\textbf{Major}\\\textbf{Release}\\ \end{tabular} 
& \begin{tabular}[t]{@{}l@{}l@{\qquad}|}\textbf{Versions}\end{tabular} 
& \begin{tabular}[t]{@{}l@{}l@{\qquad}|}\textbf{\#Classes} \end{tabular} 
& \textbf{\#Methods}  
& \textbf{\#LOC} \\
\hline

Apache Tomcat & 6 & \begin{tabular}[t]{@{}l@{}l@{\qquad}|} 6.0.16, 6.0.18, 6.0.26, 6.0.29, \\ 6.0.30, 6.0.32, 6.0.33, 6.0.35,\\ 6.0.36, 6.0.37, 6.0.39, 6.0.41, \\6.0.43 \end{tabular}& 1,763 & 16,601 & 366,948\\
\hline

& 7 & \begin{tabular}[t]{@{}l@{}l@{\qquad}|} 7.0.10, 7.0.11, 7.0.16, 7.0.20, \\7.0.21, 7.0.22, 7.0.27, 7.0.29, \\ 7.0.32, 7.0.39, 7.0.42, 7.0.47, \\ 7.0.50, 7.0.52, 7.0.53, 7.0.54, \\ 7.0.57 \end{tabular} & 1,631  & 14,859 & 325,300\\
\hline

& 8 & \begin{tabular}[t]{@{}l@{}l@{\qquad}|} 8.0.0-RC1, 8.0.0-RC5, 8.0.1, \\8.0.3, 8.0.5, 8.0.8, 8.0.15 \end{tabular}& 3,390 & 26,725 & 530,604\\
\hline
\hline

Apache CXF & 2 & \begin{tabular}[t]{@{}l@{}l@{\qquad}|} 2.5.1, 2.5.2, 2.6.0, 2.6.2, 2.7.0,\\ 2.7.2, 2.7.8, 2.7.9, 2.7.10, 2.7.11 \end{tabular}& 7,255 & 44,069 & 808,377 \\
\hline

 & 3 & \begin{tabular}[t]{@{}l@{}l@{\qquad}|} 3.0.1, 3.0.2, 3.0.3, 3.0.6, 3.0.7, \\3.1.8, 3.1.9  \end{tabular}& 8,870 & 53,545 & 964,811\\
\hline
\hline
Android & 6 & \begin{tabular}[t]{@{}l@{}l@{\qquad}|}
6.0.0.\_r41
\end{tabular}& 16,175 & 273,515 & 4,250,931\\
\hline
 & 7 & \begin{tabular}[t]{@{}l@{}l@{\qquad}|}
7.0.0.\_r34
\end{tabular}& 19,129 & 292,525 & 5,029,189\\
\hline
\hline
\textbf{Combined} &  & \begin{tabular}[t]{@{}l@{}l@{\qquad}|}
\end{tabular}& \textbf{58,213} & \textbf{721,839} & \textbf{12,276,160}\\
\hline
\end{tabular}
}
\end{table}

Apache Tomcat is a web application server developed by the Apache Software Foundation.\footnote{http://tomcat.apache.org/} The software is implemented in Java and has about a half-million lines of code in each version. Apache CXF\footnote{https://cxf.apache.org/} is an open-source service framework that helps to build and develop services using front-end programming APIs. 
For our study, we considered all the classes of Apache Tomcat that were reported as having vulnerabilities in the security page\footnote{\label{tomcat}https://tomcat.apache.org/security.html} for major release versions 6, 7, and 8. We use the term \emph{vulnerable classes} to describe these classes where vulnerabilities were reported. On the other hand, all classes in major releases 6, 7, and 8 that do not have any vulnerabilities reported in the security page are considered as \emph{neutral classes}. Similarly, for CXF, we presented all the versions considered in the study (Table~\ref{stats}).

We used the Android Open Source Platform (AOSP) repository in our analysis because Google (the company that maintains AOSP) publishes a list of recently-patched software vulnerabilities in AOSP each month.\footnote{https://source.android.com/security/bulletin/} AOSP also contains a substantial amount of Java code, especially in layers related to the user interface. We chose the base Android platform repository (/platform/frameworks/base/) because it is a large, primarily Java codebase with a high number of security patches and it is central to AOSP. We analyzed all the vulnerable and neutral classes of Android versions 6.0.0\_r41 and 7.0.0\_r34 for this study. 



\subsection{Data Extraction}

\subsubsection{Extracting Software Vulnerabilities}
For the Apache projects, we collected vulnerabilities from the Apache Tomcat\textsuperscript{\ref{tomcat}} and Apache CXF\footnote{\label{cxf}http://cxf.apache.org/security-advisories.html} security pages. The vulnerability reports provide the information about the vulnerability type, i.e. its CVE id (Common Vulnerabilities and Exposures), affected versions, revision number, fixed version, and severity level. The link with revision number points to the list of classes that were modified to fix the vulnerability. The security page also provides the affected code versions from where we collected the vulnerable classes. 
We then downloaded the code for all versions listed in Table~\ref{stats} and separated the vulnerable classes from the non-vulnerable classes. We also removed duplicated classes that occurred in multiple versions of a system. The source code for Apache Tomcat and Apache CXF is located in the Apache Archives of Tomcat\footnote{http://archive.apache.org/dist/tomcat} and CXF\footnote{http://archive.apache.org/dist/cxf/} respectively. For example, a Denial of Service (CVE-2014-0075) vulnerability was fixed in revision 1578341 of version 7.0.53. If we follow the link to the revision number\footnote{http://svn.apache.org/viewvc?view=revision\&revision=1578341}, we get the list of classes modified to fix the vulnerability. We considered 7.0.52 as an affected version and considered an affected Java class in version 7.0.52 as a vulnerable class. Any class of that version having no reported vulnerability is considered as a neutral class. \\

\begin{table*}[!htb]
\begin{minipage}{\textwidth} 
\caption{Distribution of Smells across Vulnerable and Neutral Classes}
\label{distribution}
\centering
\scalebox{0.9}{
\begin{tabular}{cc|cccc|ccc}
\hline
\textbf{Systems} & 
\textbf{Major Release} & 
\multicolumn{4}{c|}{\textbf{Vulnerable Classes}} & \multicolumn{3}{c}{\textbf{Neutral Classes}} \\ 
\hline
 &  & 
 \textbf{\begin{tabular}[c]{@{}c@{}}Classes with \\ Smells\end{tabular}} & 
 \textbf{\begin{tabular}[c]{@{}c@{}}Classes with \\ no Smells\end{tabular}} & \textbf{\begin{tabular}[c]{@{}c@{}}\#Vulnerabilities\\ (in Classes with \\Smells)\end{tabular}} & \textbf{\begin{tabular}[c]{@{}c@{}}\#Smells\\ (in Classes with \\Smells)\end{tabular}} & 
 \textbf{\begin{tabular}[c]{@{}c@{}}Classes with \\ Smells\end{tabular}} & 
 \textbf{\begin{tabular}[c]{@{}c@{}}Classes with \\ no Smells\end{tabular}} & 
 \textbf{\begin{tabular}[c]{@{}c@{}}\#Smells\\ (in Classes with \\Smells)\end{tabular}} \\
 \hline
Apache Tomcat & 6 & 54 & 0 & 124 & 1,768 & 17,921 & 177 & 154,618 \\
\hline
\textbf{} & 7 & 56 & 0 & 106 & 1,951 & 35,956 & 676 & 256,548 \\
\hline
\textbf{} & 8 & 10 & 0 & 21 & 72 & 16,209 & 420 & 110,694 \\
\hline
\textbf{} & Total & 120 & 0 & 251 & 4,042 & 70,086 & 1,273 & 523,133 \\
\hline
\hline
Apache CXF & 2 & 22 & 0 & 31 & 380 & 69,459 & 2,402 & 429,467 \\
\hline
\textbf{} & 3 & 6 & 0 & 6 & 68 & 49,179 & 2,002 & 306,191 \\
\hline
\textbf{} & Total & 28 & 0 & 37 & 448 & 118,638 & 4,404 & 735,658 \\
\hline
\hline
Android & 6 & 64 & 0 & 85 & 6,895 & 9,767 & 118 & 115,630 \\
\hline
\textbf{} & 7 & 81 & 0 & 116 & 8,705 & 11,362 & 131 & 139,489 \\
\hline
\textbf{} & Total & 145 & 0 & 201 & 15,600 & 21,129  & 249 & 255,119 \\
\hline
\end{tabular}
}
\end{minipage}
\end{table*}

\subsubsection{Extracting Smells}
\label{cs extract}
Next, we extracted the smells using a custom-built tool called GetSmells.\footnote{\label{gsurl}https://github.com/tdresearchgroup/getsmells} First, the projects' source files are parsed by the Scitool Understand Command Line tool\footnote{https://scitools.com/} to extract information on the different artifacts (classes, methods, variables, and so on). Then, GetSmells uses static metrics,  based on parameters of source code such as size, complexity, and inheritance, obtained from SciTool Understand to generate the smells, at class and method levels, according to well-established predefined rules-based detection strategies and algorithms from existing studies in the literature e.g.~\cite{Lanza, fontana2017arcan}. A PDF document (rules.pdf) describing these rules and algorithms and referring to the source of the rules is provided in the GetSmells repository\footnotemark\protect[\ref{gsurl}]. As an example, the detection rule for the God Class code smell is provided in the Background (Section~\ref{background}). GetSmells extracts 16 smells in total - 8 at the class level, 5 at the method level, 2 at the package level, and 1 at the file level. Despite extracting the smells at different levels of granularity, we considered the smells at the class level only to be able to establish a relationship with vulnerabilities that are reported at the class level. For example, if class A has a smelly method B, then we consider class A as being smelly. Similarly, if a file C is smelly and contains classes D and E, we consider classes D and E as smelly. 

Our choice of smells was influenced by several factors:
\begin{itemize}
\item {Smells with detection rules that have been defined and well-documented in the literature and also used in previous studies~\cite{Lanza,fontana2017arcan}.}
\item {Smells with a negative impact on software quality~\cite{olbrich2010, Khomh, codabux2017relationship, codabux2017}. }
\item {Inclusion of package and file-level smells which are often overlooked in smell studies in addition to class and method level smells.}
\end{itemize}

We ran GetSmells on Apache Tomcat, Apache CXF, and Android. We extracted smells for all vulnerable classes of the affected versions and neutral classes of Apache Tomcat, Apache CXF, and Android.    

\subsection{Data Analysis}
From the data extraction step, we have a list of vulnerable and neutral classes for the different versions of Apache Tomcat, Apache CXF, and Android as well as the smells for both sets of classes for each system. With all the projects combined, we analyzed about 216,000 classes. 

For RQ1, we test whether smells are associated with vulnerable and neutral classes. We use Fisher's exact test \cite{fisher1992}, which checks whether a proportion varies between two samples. Fisher's exact test is very similar to the Chi-Square test but  Fisher's exact test is used in cases when one of the four cells of a 2 x 2 contingency table has less than five observations, as is the case with the vulnerable classes with no smells. As our data is not normally distributed, we chose Fisher's exact test which is a non-parametric test~\cite{winters2010statistics}. $p-values$ (which can take any value between zero and one - 0 suggesting that the observed difference being due to chance, and 1 suggesting that there is no difference between the groups other than due to chance) are not interpreted with Fisher's exact test. Instead, the odds ratio with 95\% confidence interval is used. Therefore, in  Table~\ref{fisher}, we also compute the odds ratio (OR) \cite{fisher1992} that indicates the likelihood for an event to occur. An OR of 1 indicates that the event is equally likely in both samples. An OR higher than 1 indicates that the event is more likely in the first sample (with smells), while an OR lower than 1 indicates that it is more likely in the second sample (without smells).

For RQ2, we performed a Chi-Square Test of Independence to compare the difference in the distribution of smells across the vulnerable and neutral classes. We performed separate Chi-Square Tests of Independence for the different types of smells.
For degrees of freedom of $1$ and at 5\% level of significance, the appropriate critical value is 3.84 and the decision rule is as follows: Reject $H_0$ if $\tilde\chi^2$ $\geq$ 3.84. In Table~\ref{chi}, we also reported \textit{Yates correction} along with the chi-square values. The Yates correction is a correction that is used for both Pearson’s chi-square test and McNemar’s chi-square test to reduce their bias towards upwards for a $2$ x $2$ contingency table~\cite{yates}. 

We can reject our null hypothesis if $\tilde\chi^2$ $\geq$ 3.84 for any smell, meaning the smells are more likely to be related to software vulnerability. In other words, that particular smell has a relationship with the vulnerability as the frequency distribution of that smell in the vulnerable vs. neutral classes shows a statistically significant difference ($p-value$ is less than .05). On the other hand, if $\tilde\chi^2 < 3.84$ for any smell, we cannot reject the null hypothesis, which means there is no statistical evidence that the smell is more likely to make the class vulnerable. In other words, that particular smell may or may not have any relationship with vulnerability as the distribution of that smell in the vulnerable vs. neutral classes does not show a statistically significant difference ($p-value$ is greater than or equal to .05). 

\subsection{Replication Package}
All the data used in our study are publicly available.\footnote{https://bit.ly/2yNytx5} Specifically, we provide the Python scripts and the working data sets used to run the statistical analysis reported in this paper. We also provide the GetSmells tool and scripts that we used to extract the smells and perform data analysis and processing\footnotemark\protect[\ref{gsurl}].

\section{Results}
\label{results}
In this section, we report the results of our study to address the research questions.
\subsection{RQ1: What is the relationship between smells and software vulnerabilities?}
In Table~\ref{fisher}, we combined the data from different versions of a particular system. For each system, we report the number of (i) vulnerable classes with smells (ii) vulnerable classes with no smells (iii) non-vulnerable classes with smells (iv) non-vulnerable classes with no smells as well as the results of Fisher's exact test when testing the null hypothesis.

\begin{table*}[!htb]
\begin{minipage}{\textwidth} 
\caption{Contingency Table and Fisher's Exact Test Results}
\label{fisher}
\centering
\scalebox{0.8}{
\begin{tabular}{lccccccc}
\hline
\textbf{Systems} 
& \textbf{Release} 
& \textbf{Smells \& Vulnerability} 
& \textbf{No Smells \& Vulnerability}
& \textbf{Smells \& No Vulnerability} 
& \textbf{No Smells \& No Vulnerability} 
& \textbf{$p-value$}
& \textbf{OR}\\
\hline
Apache Tomcat & 6 & 54 & 0 & 17,921 & 177 & 1 & 1.08\\
& 7 & 56 & 0 & 35,956 & 676 & 0.63 & 2.13\\
& 8 & 10 & 0 & 16,209 & 420 & 1 & 0.54\\
\hline
& Total & 120 & 0 & 70,086 & 1,273 & 0.28 & 4.38\\
\hline
Apache CXF & 2 & 22 & 0 & 69,459 & 2,402 & 1 & 1.56\\
& 3 & 6 & 0 & 49,179 & 2,002 & 1 & 0.53\\
\hline
& Total & 28 & 0 & 118,638 & 4,404 & 0.63 & 2.12\\
\hline
Android& 6 & 64 & 0 & 9,767 & 118 & 1 & 1.57\\
& 7 & 81 & 0 & 11,362 & 131 & 1 & 1.89\\
\hline
& Total & 145 & 0 & 21,129 & 249 & 0.42 & 3.44\\
\hline
\hline
\textbf{Combined}& & \textbf{293} & \textbf{0} & \textbf{209,853} & \textbf{5,926} & \textbf{0.0005} \textbf{($<$ {.05})} & \textbf{16.58} \\
\hline

\end{tabular}
}
\end{minipage}
\end{table*}

For the individual versions of all three systems, the $p-value$ is greater than .05. Therefore, the results are not statistically significant and indicate strong evidence for the null hypothesis. However, when the systems are not differentiated and the data is combined, the $p-value$ is less than .05 indicating that the proportions are significantly different, thus allowing to reject $H_{01}$. Regarding the ORs, for Apache Tomcat 6, the odds are even when comparing whether a class with smell has a vulnerability or not. Except for Apache Tomcat 8 and Apache CXF 3, the odds of a class that has a vulnerability are more favorable that it also contains a smell. When the systems are not differentiated, the OR is greater than 1 confirming the above --- that there are more favorable odds for a class that has a vulnerability to also have a smell. So, we conclude that the odds for a class to be vulnerable are higher for classes with smells. 

\begin{tcolorbox}
\textbf{RQ1 Summary:} We found that classes with smells are more likely to be vulnerable. 
\end{tcolorbox}

\begin{table}[!htb]
\begin{minipage}{\linewidth}
\caption{Chi-Square Statistics - Smells}
\label{chi}
\centering
\resizebox{\linewidth}{!}{
\begin{tabular}{lccc}
\hline
\textbf{Systems} & \textbf{Apache Tomcat} & \textbf{Apache CXF} & \textbf{Android} \\
\hline
God Class & 334.63 (325.97) 
& 52.26 (45.09) & 628.06 (604.05)\\

Lazy Class & - & - & 202.46 (198.87)\\	

Complex Class & 337.67 (332.73)
& 126.19 (117.51) & 308.73 (303.83)\\  

Large Class	& 225.16 (221.99) & 11.41 (9.99) & 63.28 (61.79)\\

Refused Bequest & 20.14 (18.46)
& 11.11 (7.82) & 32.32 (29.93)\\

Data Class	& 49.15 (47.88)
& 8.66 (7.58) & 22.23 (21.44)\\

Feature Envy & 196.19 (192.91) & 8.66 (7.1) & 252.29 (248.95)\\

Brain Class & 319.64 (280.01) & 149.64 (36.91) & 10.19 (5.10)\\

Hub-Like Dependency & 3.85 \textcolor{red} {(3.04)} & \textcolor{red}{3.45 (1.88)} &	9.56(8.49)\\

Class Cyclic Dependency & \textcolor{red} { 0.003 (.009)} & - & 53.26 (51.39)\\  

Unhealthy Inheritance Hierarchy & 13.87 (12.36) & \textcolor{red}{3.83 (2.58)} & 121.82 (117.63)\\

Long Method & 62.93 (61.44) & - & 25.83 (24.84)\\

Long\_Parameter\_List  & 42.04 (40.81) & 19.53 (17.89) & 19.84 (18.91)
\\

Shotgun Surgery & 4.43 \textcolor{red} {(3.73)} & \textcolor{red}{2.02 (1.12)} & 185.99 (181.83)\\	

Brain Method & 280.58 (268.09)
 & 6.28 \textcolor{red}{(1.16)} & 159.39 (148.05)\\

Unstable Dependency & \textcolor{red}{0.48 (0.30)} & \textcolor{red}{0.66 (0.23)} & 
\textcolor{red}{0.21 (0.13)}
\\ 

Package Cyclic Dependency & \textcolor{red}{0.33 (0.19)} & \textcolor{red}{0.39 (0.07)} & 
\textcolor{red}{0.84 (0.51)}\\
\hline

\end{tabular}

}
\end{minipage}
\footnotesize{Note: The chi-square values mentioned within the parentheses are the values after Yates correction. For the value in black, the p-value is less than .05 (the metric shows a statistically significant difference between vulnerable and neutral classes). For the value in red, the p-value is greater than .05 (the metric does not show a statistically significant difference between vulnerable and neutral classes).}
\end{table}

\subsection{RQ2: Are there certain smells that have a stronger relationship to vulnerabilities than others?}

\subsubsection{Apache Tomcat}
\label{results-tomcat}
\hfill\\
To summarize, there are 3,791 smell instances in 120 vulnerable classes vs. 521,860 smell instances in 70,068 neutral classes. A vulnerable class has on average 32 smells while a neutral class has on average 7 smells. 

The top two prominent smells in both vulnerable and neutral classes are \textit{Long Parameter List} and \textit{Long Method}. There is a significant difference in the distribution of these smells in the vulnerable and neutral classes. According to Table~\ref{chi}, for Apache Tomcat, the following smells have chi-square values greater than $3.84$ after Yates correction: \textit{God Class, Complex Class, Large Class, Refused Bequest, Data Class, Feature Envy, Brain Class, Unhealthy Inheritance Hierarchy, Long Method, Long Parameter List,} and \textit{Brain Method}. For these smells, the chi-square test shows a statistically significant difference in vulnerable and neutral classes, and therefore, for those smells, we can reject the null hypothesis. Five smells have chi-square values less than $3.84$ after Yates correction: \textit{Hub-Like Dependency, Class Cyclic Dependency, Shotgun Surgery, Unstable Dependency,} and \textit{Package Cyclic Dependency}. For these smells, the chi-square test does not show a statistically significant difference in vulnerable and neutral classes and therefore, for those smells, we cannot reject the null hypothesis. In other words, we cannot claim that a class containing those types of smells is more likely to have a software vulnerability. We could not conduct the chi-square test for \textit{Lazy Class} smell as it does not occur in the dataset. 

\subsubsection{Apache CXF}
\hfill\\
To summarize, there are 448 smell instances in 28 vulnerable classes versus 735,658 smell instances in 118,638 neutral classes. A vulnerable class has on average 15 smells while 1 in 15 neutral classes have a smell. 
Again, the top two prominent smells in both vulnerable and neutral classes are \textit{Long Parameter List and Long Method}. There is a significant difference in the distribution of smells in the vulnerable and neutral classes. According to Table~\ref{chi}, for Apache CXF, the following smells have chi-square values greater than $3.84$ after Yates correction: \textit{God Class, Complex Class, Large Class, Refused Bequest, Data Class, Feature Envy, Brain Class and Long Parameter List}. For these metrics, the chi-square test shows a statistically significant difference in vulnerable and neutral classes, therefore, we can reject the null hypothesis. Six smells have chi-square values less than $3.84$ after Yates correction: \textit{Hub-Like Dependency, Unhealthy Inheritance Hierarchy, Shotgun Surgery, Brain Method, Unstable Dependency, and Package Cyclic Dependency}. For these smells, the chi-square test does not show a statistically significant difference in vulnerable and neutral classes, therefore, we cannot reject the null hypothesis. In other words, we cannot claim that a class containing those types of smells is more likely to have a software vulnerability. We could not conduct the chi-square test for \textit{Lazy Class, Class Cyclic Dependency, and Long Method} smells as they do not occur in the dataset. 

\subsubsection{Android}
\hfill\\
To summarize, there are 15,600 smell instances in 145 vulnerable classes versus 255,119 smell instances in 21,129 neutral classes. A vulnerable class has on average 94 smells while 1 in 13 neutral classes have a smell. 

As with the other two projects, the top two smells in both vulnerable and neutral classes are \textit{Long Parameter List and Long Method}. There is a significant difference in the distribution of smells in the vulnerable and neutral classes. According to Table~\ref{chi}, for Android, the following smells have chi-square values greater than $3.84$ after Yates correction: \textit{God Class, Lazy Class, Complex Class, Large Class, Refused Bequest, Data Class, Feature Envy, Brain Class, Hub-Like Dependency, Class Cyclic Dependency, Unhealthy Inheritance Hierarchy, Long Method, Long Parameter List, Shotgun Surgery and Brain Method}. For these smells, the chi-square test shows a statistically significant difference in vulnerable and neutral classes, and therefore, for those smells, we can reject the null hypothesis. Two smells \textit{Unstable Dependency and Package Cyclic Dependency} have chi-square values less than $3.84$. For other smells, the chi-square test shows a statistically significant difference in vulnerable and neutral classes, therefore, for those smells, we can reject the null hypothesis. In other words, we can claim that a class containing those smells (having a statistically significant difference in the vulnerable vs. neutral classes) is more likely to contain a software vulnerability.

\begin{tcolorbox}
\textbf{RQ2 Summary:} For the systems studied, the smells which show a statistically significant relationship with software vulnerabilities are: \textit{God Class, Complex Class, Large Class, Data Class, Refused Bequest, Feature Envy, Long Parameter List and Brain Class.}
\end{tcolorbox}

\section{Discussion}
\label{discuss}

In this section, we discuss our results and present the reasoning behind each finding in our empirical analysis.
\subsection{Smells Having Significant Relation with Vulnerability}
As in Table~\ref{chi}, some code smells (\textit{God Class, Complex Class, Large Class, Data Class, Refused Bequest, Feature Envy, Long Parameter List, and Brain Class}) have a statistically significant relationship with vulnerability in each of the systems under study. These results show that the presence of these smells in classes potentially indicates that the class is vulnerable.

We see that \textit{God Class, Complex Class, Large Class} are the prominent smells in vulnerable classes (e.g. 36\% of total vulnerable classes in Apache Tomcat are \textit{God Class}).
A \textit{God Class} is a class that does more than it should, and there is a high probability that this class is a complex class too. Researchers previously found how complex code is related to vulnerabilities and how different complexity metrics can be used to predict them~\cite{YShin, YShin2}. According to the study by Cairo et al.~\cite{Cairo2018}, \textit{God Class} is a significant contributor and is positively associated with error proneness. Olbrich et al. showed that instances of \textit{God Class} and \textit{Brain Class} suffer more frequent changes and contain more defects than classes not affected by those smells when the class size is not considered~\cite{olbrich2010}. As \textit{God Class, Complex Class, and Large Class} are structurally similar to each other, they have the potential to introduce defects and vulnerabilities. 

Gradisnik et al.~\cite{Gradisnik2018} showed that \textit{Feature Envy} is expected to be correlated with low quality code. Feature envy occurs when one object depends on another object for some computations and exposes its data fields to another object instead of doing the computation itself. For example, consider a class Rectangle that exposes its width and height fields to the other classes for computing its area. \textit{Feature Envy} hampers encapsulation and can be responsible for leaking information. In Apache Tomcat, we found this smell in 68.3\% of the total vulnerable classes.

\textit{Shotgun Surgery} has been found as a significant contributor in Eclipse 2.1 and positively associated with change proneness and error proneness at all severity levels~\cite{Cairo2018}. \textit{Shotgun Surgery} occurs when a single responsibility has been split up among a large number of classes. It makes the code extremely non-cohesive and localization of similar changes becomes harder which makes the code vulnerable. In our study, \textit{Shotgun Surgery} exhibited a significant relationship with vulnerabilities in both Apache Tomcat and Android. 

We also found a significant relationship between \textit{Long Method} and \textit{Brain Method} smells with vulnerability as shown in Table~\ref{chi}. 
\textit{Long Method} represents a single method and characterizes situations in which the method is excessively long, making it difficult to understand or (re)use~\cite{Cairo2018}. The presence of \textit{Long Method} might make maintenance tasks difficult in terms of effort and time which will increase the possibility of defects~\cite{Chatzigeorgiou}. A \textit{Brain Method} centralizes the intelligence of a class and manifests itself as a long and complex method that is difficult to be understood and maintained by the developers~\cite{Vidal}. 

There was no relevant relationship between \textit{Data Class} and \textit{Refused Bequest} and the occurrence of bugs in a prior study by Cairo et al.~\cite{Cairo2018}. Our study, however, did find a statistically significant relationship between \textit{Data Class} and \textit{Refused Bequest} with software vulnerabilities. This result should be investigated further as the study by Cairo et al. could not correlate the two smells with more general bugs.

\textbf{In summary, we found that the smells \textit{God Class, Complex Class, Large Class, Feature Envy, Long Parameter List} and \textit{Brain Class} are correlated with software vulnerabilities across all systems studied.
Existing studies support our findings by identifying these smells as potential indicators of other issues (e.g., defects). Besides, we also found a statistically significant relationship between \textit{Data Class and Refused Bequest} and software vulnerabilities in contrast to the study by Cairo et al.~\cite{Cairo2018} which could not establish any relationship between these smells and defects.} 

\subsection{Smells Having No Significant Relation with Vulnerability}
 We see in Table~\ref{chi} that architecture smells including \textit{Hub-Like Dependency, Class Cyclic Dependency, Unstable Dependency, Package Cyclic Dependency} do not exhibit a statistically significant difference between vulnerable and neutral classes. These smells indicate a tendency for code to contain dependencies (cyclic and otherwise) at the architectural level. According to Fontana et al., the architectural subsystems involved in a dependency cycle are hard to release, maintain, and reuse. The authors encouraged further study on the potential correlations between architectural smells and bugs to improve the accuracy of bug prediction~\cite{Fontana2019}. Hence, earlier study by Fontana et al.~\cite{Fontana2019} partially motivates this study, although we did not find a relationship between architectural smells and the vulnerabilities in the systems analyzed.

Mo et al.~\cite{Mo} showed that the files involved in architectural patterns have significantly higher bugs and change rates compared to the average files in a project. It is not surprising that historical and structural dependencies among these files may incur architectural flaws which might lead to software vulnerabilities~\cite{Feng}. Earlier works~\cite{Mo, Feng} focused mostly on design decisions that result in architectural flaws called \textit{hotspot patterns}. Hotspot patterns are the direct violation of design rules whereas the architecture smells considered here are non-volitional design weaknesses which may relate to deeper architectural problems.  

\textbf{In summary, we found no significant relationship between architectural smells and software vulnerabilities. Despite the significant relationship between certain code smells and software vulnerability, there was no corresponding relationship with architecture smells.}

\subsection{How Our Findings Differ From Others?}
As elaborated in Section~\ref{results}, we found that smells have a relationship with software vulnerabilities. This is supported by Eklkhail et al.~\cite{Elkhail}, Mumtaz et al.~\cite{Mumtaz}, Ghafari et al.~\cite{Ghafari}, Rahman et al.~\cite{rahman2019, rahman2019share}.

However, Eklkhail et al.~\cite{Elkhail} found that "Field Declaration Not Start At Class" code smell is the most correlated code smell in vulnerable classes whereas our study identifies \textit{God Class, Complex Class, Large Class, Data Class, Feature Envy, Brain Class} as being the most correlated with software vulnerabilities. We know that \textit{God Class} is a class that does more than it is supposed to do. Although it is highly likely that \textit{God Class, Large Class and Complex Class} have similar characteristics, they are still distinct smells. For example, a class with thousands lines of code can be a large class, while it might not be a complex class if it has a small number of decision statements or a small number of independent paths. Therefore, there is a need to study them separately.

Mumtaz et al.~\cite{Mumtaz} reported \textit{Feature Envy} and \textit{Data Class} code smells to be the most prominent code smells in vulnerable classes. In our study, we found \textit{Long Parameter List and Long Method} code smells to be the most popular smells in vulnerable classes. Ghafari et al.~\cite{Ghafari} and Gadient~\cite{gadient2019} identified security code smells that lead to vulnerabilities in Android apps based on a literature survey. In~\cite{gadient2019}, the authors used a static analysis tool to extract the security smells and considered Inter-Component Communication (ICC) vulnerabilities in Android. In our study, we focused on publicly available vulnerability reports for realistic vulnerability data, making our data more reliable and our study more robust (i.e., we used vulnerabilities that were reported as CVEs or confirmed by the project owners). 

Rahman et al.~\cite{rahman2019} conducted an empirical study focused on security smells in IaC scripts. They found that security smells existed across all datasets they investigated but the hard-coded secret was the most prominent security smell out of the smells they investigated. In another study, Rahman et al.~\cite{rahman2019share} conducted a similar study on GitHub Python Gists and found command injection to be the most prevalent security smell. The smells considered in this study are different than the smells in~\cite{rahman2019share}. They focused on security smells which may directly relate to vulnerabilities by design. On the other hand, we investigate traditional code and architectural smells and show how they can be related to the vulnerable code. Revealing the relationship between traditional code smells and vulnerabilities will guide developers during the development process as those smells more frequently occur while developing the code. The goal is eventually tool integration to enable more secure coding practice and thus mitigating certain vulnerabilities that may exist in smelly code.

The study on the relationship between architectural smells and vulnerabilities is also significant in software security research. Although we did not find any statistically significant relationship between them, with a strong relationship between certain code smells and vulnerability, we posit that a more thorough investigation of architectural smells is warranted.  

\textbf{Some of the studies above target specific domains e.g Android apps, IaC scripts, or GitHub Python Gists for their experiments. Several studies also focus on security smells while our study focuses on predominant smells that are the logical consequences of writing code and enable complex features and robust user interactions. While we cannot perform a direct comparison with some of the existing works as to which smells are the most prominent, the prior research does corroborate our finding that certain smells are related to software vulnerabilities.}


\subsection{Relevance to Practitioners}
We envision the tools and methodology described as plugins attached to developer IDEs and continuous integration (CI) workflows. The results show that certain smells are more prone to vulnerability than others. Whether it be testers ensuring the exhibited smells are covered with targeted tests, code reviews conducted where the module owner must recheck code with certain smells, or a rejection of certain smelly code that falls within a metric threshold by the CI server, these results provide a means to further improve code security and can be easily integrated into existing tools and workflows.

\section{Threats to validity}
\label{threats}
Below, we identify threats to the validity of the research.  

\noindent \textbf{Construct Validity}
We extracted smells using the custom-built tool GetSmells. We chose the set of smells based on the factors mentioned in Section~\ref{cs extract}. Not considering other code and architecture smells and the efficiency of the tool used may limit the validity of our work. In our study, we collected only the reported vulnerabilities. We considered all the classes in the projects' major releases that do not have any reported vulnerabilities as neutral classes in this study. We understand that there could be hidden or unreported vulnerabilities in those classes that were not taken into consideration.

\noindent \textbf{External Validity} 
The experiments were conducted for software implemented in the Java programming language. Therefore, our results cannot be generalized to other systems written in different, non object-oriented programming languages. Moreover, we only considered the reported vulnerabilities. The vulnerabilities which have not been released or reported for the systems studied have not been considered. Besides, despite two of the analyzed projects being from the Apache family, we ensured that the teams consist of different people.\footnote{http://tomcat.apache.org/whoweare.html}\footnote{https://cxf.apache.org/people.html} 

\noindent \textbf{Internal Validity} 
Some confounding factors (e.g. size of the classes, number of instances of code smells per class) which have not been considered in our study might have an impact on our results. We are not claiming causation, just associating software vulnerabilities with the presence of smells. In Section~\ref{discuss}, we discussed the possible reasons why some smells are proportionally related to software vulnerabilities.


\section{Conclusion}
\label{conc}

In this paper, we present a study evaluating three major releases of Apache Tomcat, two versions of Apache CXF, and two versions of Android to determine the relationship between smells and software vulnerabilities. Compared to previous studies which used static analysis tools to identify vulnerabilities, incurring risks of a large number of false positives which might impact their results or studies that focused on a particular type of code smell (e.g. duplicated code), our study considered only reported vulnerabilities which eliminates the occurrence of false positives. 
We empirically found that certain smells exhibit a statistically significant difference in vulnerable vs. neutral classes whereas others do not show significant differences. This result indicates that there is a relationship between smells (that show significant differences in two groups: vulnerable and neutral classes) and software vulnerabilities. We know that a code smell is a surface indication of a deeper problem in code. Identifying the relationship between smells and vulnerabilities can help developers identify deeper problems in their code and address them before deployment. The objective is to mitigate the risks of releasing vulnerable code.

This study provides a basis for future work to determine the underlying reason behind the different distributions of the smells in vulnerable and neutral classes. In this study, we did not make any distinction between different types of smells or the complexity of the different classes or the quantity of the distinct smells. We also did not consider the types of vulnerabilities relevant to smells. These can be explored further in the future studies. We also plan to extend the study to other systems as well as use these findings to create software vulnerability prediction models.

\section{Acknowledgement}
\label{ack}
The authors would like to acknowledge Yuhan Hu (now at Calian SED Systems, Canada) and Charles Boyd (now at Google, USA) who helped build GetSmells and extract vulnerabilities. Hu implemented additional tools to facilitate the data analysis and processing as well.

\bibliographystyle{IEEEtran}
\bibliography{Apsec}

\end{document}